\documentclass[aps,twocolumn,superscriptaddress,showpacs]{revtex4}

\usepackage[dvips]{graphicx}
\newcommand{\nc}{\newcommand}           
\nc{\vc}[1]     {\mbox{\boldmath $#1$}} 
\nc{\wtil}      {\widetilde}            

\def\JL#1#2#3#4{ {{\rm #1}} \textbf{#2}, #4 (#3)}  
\nc{\PR}[3]     {\JL{Phys. Rev.}{#1}{#2}{#3}}
\nc{\PRC}[3]    {\JL{Phys. Rev.~C}{#1}{#2}{#3}}
\nc{\PRA}[3]    {\JL{Phys. Rev.~A}{#1}{#2}{#3}}
\nc{\PRL}[3]    {\JL{Phys. Rev. Lett.}{#1}{#2}{#3}}
\nc{\NP}[3]     {\JL{Nucl. Phys.}{#1}{#2}{#3}}
\nc{\NPA}[3]    {\JL{Nucl. Phys.}{A#1}{#2}{#3}}
\nc{\PL}[3]     {\JL{Phys. Lett.}{#1}{#2}{#3}}
\nc{\PLB}[3]    {\JL{Phys. Lett.~B}{#1}{#2}{#3}}
\nc{\PTP}[3]    {\JL{Prog. Theor. Phys.}{#1}{#2}{#3}}
\nc{\PTPS}[3]   {\JL{Prog. Theor. Phys. Suppl.}{#1}{#2}{#3}}
\nc{\PRep}[3]   {\JL{Phys. Rep.}{#1}{#2}{#3}}
\nc{\JP}[3]     {\JL{J. of Phys.}{#1}{#2}{#3}}
\nc{\andvol}[3] {{\it ibid.}\JL{}{#1}{#2}{#3}}

\begin{document}

\title{
Resonances of $^7$He in the complex scaling method
}

\author{Takayuki Myo
\footnote{E-mail address: myo@rcnp.osaka-u.ac.jp}
}
\affiliation{
Research Center for Nuclear Physics (RCNP),
Ibaraki, Osaka 567-0047, Japan}

\author{Kiyoshi Kat\=o
\footnote{E-mail address: kato@nucl.sci.hokudai.ac.jp}
}
\affiliation{
Division of Physics, Graduate School of Science,
Hokkaido University, Sapporo 060-0810, Japan}

\author{Kiyomi Ikeda
\footnote{E-mail address: k-ikeda@postman.riken.go.jp}
}
\affiliation{
RIKEN Nishina Center, 2-1 Hirosawa, Wako, Saitama 351-0198, Japan}

\date{\today}

\begin{abstract}
We study the resonance spectroscopy of $^7$He in the $^4$He+$n$+$n$+$n$ cluster model,
where the motion of valence neutrons is described in the cluster orbital shell model.
Many-body resonances are treated on the correct boundary condition as the Gamow states in the complex scaling method.
We obtain five resonances and investigate their properties from the configurations.
In particular, the $1/2^-$ state is found in a low excitation energy of 1.1 MeV with a width of 2.2 MeV,
while the experimental determination of the position of this state is not so clear.
We also evaluate the spectroscopic factors of the $^6$He-$n$ components in the obtained $^7$He resonances. The importance of the $^6$He($2^+$) state is shown in several states of $^7$He.
\end{abstract}



\maketitle 

\section{Introduction}

Development of the radioactive beam experiments provides us with much information of the unstable nuclei far from the stability.
In particular, the light nuclei near the neutron drip-line exhibit the new phenomena of the nuclear structures,
such as the neutron halo structure in $^6$He, $^{11}$Li and so on.
The disappearance of the $0p$-$1s$ shell gap is also found in $^{11}$Li and neighboring nuclei\cite{Ta92,Ta96}.

Recently, many experiments of $^7$He, the unbound nuclei, have been reported\cite{Ko99,Bo01,Me02,Wu05,Bo05,Sk06,Ry06,Be07}.
The ground state is commonly assigned to be the $3/2^-$ resonant state at 0.3-0.5 MeV above the $^6$He+$n$ threshold energy.
However, there are still found contradictions in the observed energy levels and 
the excited states are not settled for their spins and energies. 
The excited state at $E_x \sim 3$ MeV is reported in several experiments\cite{Ko99,Bo01,Wu05,Ry06} and 
a possibility of the $5/2^-$ state is proposed in Refs.\cite{Sk06,Ry06}. 
The existence of $1/2^-$ and $3/2^-_2$ states is also expected\cite{Me02,Bo05,Wu05,Sk06,Ry06}, but still unclear and their positions 
and decay widths are not fixed.
In particular, the $1/2^-$ state is interested with the possibility of the $LS$ partner of the ground $3/2^-$ state,
because the $LS$ splitting in this nucleus may give an important information on the $LS$ interaction in neutron drip-line nuclei.
For this state, the recent experiments\cite{Me02,Sk06,Ry06} report it with the low excitation energy at around the 1 MeV region. 
On the other hand, other observations\cite{Wu05,Bo05} exclude the low excitation energy of $1/2^-$ reported in Ref.~\cite{Me02}
and suggest a little higher excitation energy\cite{Wu05}.

In the theoretical side, {\it ab initio} calculations of the no-core shell model\cite{Na98} and the Green's function Monte Calro\cite{Pi04} have been performed, and the calculated energy positions of the ground state and the $5/2^-$ state show a good correspondence with the experiments\cite{Sk06}.
The $1/2^-$ state is predicted at around 3 MeV, although the theoretical results somewhat depend on the choice of the three-nucleon forces\cite{Pi04}. 
Those calculations are based on the bound state approximation and the continuum effect from many-body open channels is not taken into account correctly, though all states of $^7$He are unbound. The excited states with a few MeV excitation energy can decay not only to the two-body $^6$He+$n$ channel but also to many-body channels of $^5$He+2$n$ and $^4$He+3$n$.

Several promising methods have been proposed to take into account the continuum effects explicitly.
Starting from the traditional shell model, the particle decay into the open channels has been recently considered 
based on the continuum shell model\cite{Vo03} and the application to the He isotopes is done\cite{Vo05}.
Another approach, the so-called Gamow shell model\cite{Ha05,Ro06,Mi07}, has been presented to describe single-particle decaying states.
As for the model space, both the continuum shell model and the Gamow shell model calculations for the resonant spectroscopy of He isotopes have been carried out within $p$-shell configurations. 
It is known, on the other hand, that for the description of the weakly bound system, in addition to the $p$-shell configurations, the contributions from the higher partial waves cannot be ignored such as due to the pairing correlation. 
In particular, the $sd$-shell plays an important role and is found to give an about 1 MeV energy contribution on the binding energy of $^6$He with the appropriate interactions\cite{Ao95,My01}. 
For the spectroscopy of $^7$He, its ground state may be a single particle resonance with a $^6$He+$n$ configuration, but all other excited states are experimentally suggested to appear as two or three particle resonances above the $^4$He+3$n$ threshold energy, 
because $^6$He is a Borromean nucleus and breaks up easily into $^4$He+$n$+$n$.
Furthermore, when we discuss the properties of the $^7$He resonances, it is important to reproduce the threshold energies of the particle decays, in which the subsystems also have their particular decay widths such as $^5$He+$2n$ channels. 
This condition was not emphasized so far in the previous theoretical studies of $^7$He. Therefore, the $^{7}$He resonant spectroscopy is desired to be investigated with the appropriate treatments of the decay properties concerned with the subsystem of $^{5,6}$He, simultaneously.

The purpose of this paper is to carry out the resonance spectroscopy of $^7$He with the simultaneous descriptions of $^{5,6}$He imposing the accurate boundary conditions of many-body decays.
To do this, we employ the cluster orbital shell model (COSM) of $^4$He+$n$+$n$+$n$\cite{Su88,Ma06,Ma07}, in which the open channel effects for the $^{6}$He+$n$, $^{5}$He+2$n$ and $^{4}$He+3$n$ decays are taken into account explicitly. 
We describe the many-body resonances under the correct boundary conditions for these decay channels using the complex scaling method (CSM)\cite{ABC}.
As the details of this method are given in Ref. \cite{Ao06}, the resonant energies and decay widths of many-body resonances are directly obtained by diagonalization of the complex-scaled Hamiltonian with $L^2$ basis functions\cite{Ho83,Mo98}. 
It has been also shown that CSM is a very successful method to investigate the resonances and the Coulomb breakups of He and Li isotopes\cite{Ao95,My01,My03}.
In this paper, we find out the resonance structure of $^7$He with CSM,
and also determine the spectroscopic factors ($S$-factor) of $^6$He-$n$ components for every $^7$He resonances. The results of the $S$-factor are shown to help for understanding the coupling between $^6$He and the additional neutron in $^7$He.

\section{Complex-scaled $^4$He+$Xn$ COSM for He isotopes}

\subsection{Cluster orbital shell model (COSM) for the $^4$He+$Xn$ systems}

We explain COSM for the $^4$He+$Xn$ systems, where $X=1$ for $^5$He, $X=2$ for $^6$He and $X=3$ for $^7$He.
The Hamiltonian is the same as that used in Refs.~\cite{My01,Ma07};
\begin{eqnarray}
	H
&=&	\sum_{i=1}^{X+1}{t_i} - T_G +	\sum_{i=1}^{X}V^{\alpha n}_i + \sum_{i<j}^X V^{nn}_{ij},
        \label{eq:Ham}
\end{eqnarray}
where $t_i$ and $T_G$ are kinetic energies of each particle ($Xn$ and $^4$He) and the center of mass (cm)
of the total system, respectively.
The interactions $V^{\alpha n}$ and $V^{nn}$ are given by the so-called modified KKNN potential\cite{Ao95} for $^4$He-$n$ and 
the Minnesota potential\cite{Ta78} with 0.95 of the $u$-parameter for $n$-$n$, respectively. 
They reproduce the low-energy scattering data of the $^4$He-$n$ and the $n$-$n$ systems, respectively, which have no bound states.

\begin{figure}[t]
\centering
\includegraphics[width=8.2cm,clip]{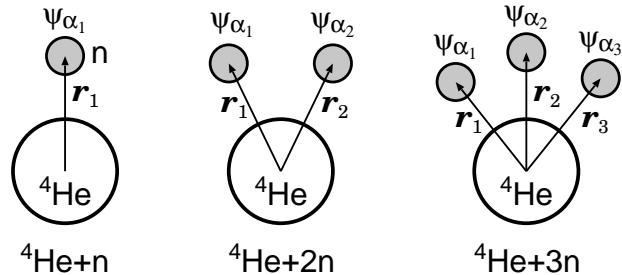}
\caption{Sets of the spatial coordinates in COSM for the $^4$He+$Xn$ system.}
\label{fig:COSM}
\end{figure}

For the wave function, $^4$He is assumed as the $(0s)^4$ configuration of a harmonic oscillator wave function, 
whose length parameter $b_c$ is taken to be 1.4 fm to fit the charge radius of $^4$He.
The motion of valence neutrons surrounding $^4$He is solved accurately using the few-body technique.
We employ a variational approach in which the relative wave functions of the $^4$He+$Xn$ system
are expanded on the COSM basis states\cite{Su88,Ma06}.
The total wave function $\Psi$ of the $^4$He+$Xn$ system is given by the superposition of 
the configuration $\Psi_\beta$ as
\begin{eqnarray}
    \Psi(^{4}{\rm He}+Xn)
&=& \sum_{\beta} C_\beta\ \Psi_\beta(^{4}{\rm He}+Xn),
    \label{WF0}
    \\
    \Psi_\beta(^{4}{\rm He}+Xn)
&=& \prod_{i=1}^X a^\dagger_{\alpha_i}|0\rangle, 
    \label{WF1}
\end{eqnarray}
where the $^4$He core is treated as a vacuum.
$a^\dagger_{\alpha_i}$ is the creation operator of the valence neutron above the $^4$He core,
with the quantum number $\alpha_i$ in a $jj$ coupling scheme. 
Here $i=1,2,3$ for three valence neutrons.
$\beta$ indicates the set of $\alpha_i$.
$C_{\beta}$ is the variational coefficient for each configuration $\Psi_\beta$ distinguished by $\beta$.
We take a summation over the available configurations.
The coordinate representation of the single particle state corresponding to $a^\dagger_{\alpha_i}$ is given as $\psi_{\alpha_i}$ with the relative coordinate $\vc{r}_i$ between the cm position of $^4$He and a valence neutron 
shown in Fig.\ref{fig:COSM}.
Including the angular momentum coupling, the total wave function $\Psi^J$ with the spin $J$ is also expressed as
\begin{eqnarray}
    \Psi^J(^{4}{\rm He}+Xn)
&=& \sum_{\beta} C_\beta\ \Psi^J_\beta(^{4}{\rm He}+Xn),
    \label{WF3}
    \\
    \Psi^J_\beta(^{4}{\rm He}+Xn)
&=& {\cal A}^\prime \left\{\, [\Phi(^4{\rm He}), \chi^{J}_\beta(Xn)]^J\, \right\},
    \label{eq:WF}
    \\
    \chi^{J}_\beta(n)
&=& \psi_{\alpha_1}^J,
    \\
    \chi^{J}_\beta(2n)
&=& {\cal A}\{ [\psi_{\alpha_1},\psi_{\alpha_2}]_J \},
    \label{eq:WF6}
    \\
    \chi^{J}_\beta(3n)
&=& {\cal A}\{ [[\psi_{\alpha_1},\psi_{\alpha_2}]_{j_{12}},\psi_{\alpha_3}]_J \}.
    \label{eq:WF7}
\end{eqnarray}
Here, as shown in Fig. \ref{fig:COSM}, $\chi^J_\beta(Xn)$ expresses the COSM wave functions for the valence neutrons.
$j_{12}$ is the coupled angular momentum of the first and second valence neutrons, 
which is included in the index $\beta$.
The antisymmetrizers between valence neutrons and
between a valence neutron and neutrons in $^4$He are expressed as ${\cal A}$ and ${\cal A}^\prime$, respectively.
The latter effect of ${\cal A}^\prime$ is treated in the orthogonality condition model\cite{My01,Ma07,Ao06}, in which
$\psi_{\alpha}$ is imposed to be orthogonal to the $0s$ state occupied by neutrons in $^4$He.
The radial part of $\psi_\alpha$ is expanded with a finite number of Gaussian basis functions\cite{Ma07} as
\begin{eqnarray}
    \psi_\alpha
&=& \sum_{k=1}^{N_\alpha} C_{\alpha,k}\ \phi_{\alpha}^k(\vc{r},b_{\alpha,k}),
    \\
    \phi_{\alpha}^k(\vc{r},b_{\alpha,k})
&=& {\cal N}\ r^{\ell_\alpha} e^{-(r/b_{\alpha,k})^2/2} [Y_{\ell_\alpha}(\hat{\vc{r}}),\chi^\sigma_{1/2}]_{j_\alpha}.
\end{eqnarray}
Here $k$ is an index for the Gaussian basis with the length parameter $b_{\alpha,k}$.
A basis number for the state $\alpha$ and the normalization factor for the basis
are given by $N_\alpha$ and ${\cal N}$, respectively. 
The expansion coefficients $\{C_\beta\}$ and $\{C_{\alpha,k}\}$ are determined variationally for the total wave function $\Psi^J$.
The length parameters $b_{\alpha,k}$ are chosen as geometric progression\cite{Hi03}.
We use at most 17 Gaussian basis functions with the max length parameter corresponding to 40 fm.

For the single particle states $\alpha=\ell_j$ $(j=\ell\otimes\frac12)$, we take angular momenta $\ell\le 2$ (up to $d$ waves) to keep the converged energy accuracy within 0.3 MeV. Namely, when we employ angular momentum states higher than  $\ell=2$, we obtain a little energy gain less than 0.3 MeV for the ground state of $^6$He\cite{Ao95}. In calculation of $^7$He, we can easily adjust the calculational energies of $^6$He by taking the 178.8 MeV of the repulsive strength of the Minnesota force\cite{Ta78} and the three-cluster interaction $V^{\alpha n n}$ for the $^4$He-$n$-$n$ system\cite{My01}. The former adjustment of the $NN$ interaction can be understood from the pairing correlation between valence neutrons with higher angular momenta $\ell>2$\cite{Ao95}. The latter is considered to come from dominantly the tensor correlation in the $^4$He core. Recently, we showed that the binding energy and excited states of $^6$He can be well explained without the three-body cluster interaction by taking into account the tensor correlation of $^4$He explicitly\cite{My05,My05b}. Here, following the previous study\cite{My01}, we use the three-cluster potential:
\begin{eqnarray}
	V^{\alpha n n}
=	\sum_{i<j} v_3\ e^{-({\bf r}_i^2+{\bf r}_j^2)/b_c^2}\,\quad \mbox{with}~~
	v_3
=	-25~{\rm MeV}.
\end{eqnarray}
Adding this three-cluster potential to the Hamiltonian in Eq.~(\ref{eq:Ham}), we obtain the observed energies of $^6$He as $-0.974$ MeV for $0^+$ and ($E_r$, $\Gamma$)=(0.840, 0.107) for $2^+$ in MeV, respectively, measured from the $^4$He+$n$+$n$ threshold.
The present model reproduces the observed energies and decay widths of $^{5,6}$He, simultaneously\cite{Aj89}, 
as shown in Fig.~\ref{fig:567}, namely, the threshold energies of the particle emissions for $^7$He.

\subsection{Complex scaling method (CSM)}
We explain CSM to obtain resonances.
In CSM, we transform the coordinates for the relative motions of the $^4$He+$Xn$ model shown in Fig.~\ref{fig:COSM}, as
\begin{eqnarray}
	\vc{r}_i
&\to&	\vc{r}_i\, e^{i\theta}
	\qquad \mbox{for}~~i=1,\cdots,X \ ,
\end{eqnarray}
where $\theta$ is the so-called scaling angle.
Using this transformation, the Hamiltonian in Eq.~(\ref{eq:Ham}) is transformed into the complex-scaled Hamiltonian $H_\theta$,
and the corresponding complex-scaled Schr\"odinger equation is given as
\begin{eqnarray}
	H_\theta\Psi^J_\theta
&=&     E\Psi^J_\theta,
	\label{eq:eigen}
	\\
        \Psi^J_\theta
&=&	e^{(3/2)i\theta\cdot X}\,
	\Psi^J(\{\vc{r}_i e^{i\theta}\}),
\end{eqnarray}
where, $X=1,2,3$ representing the number of degrees of freedom of the system.
The eigenstates are obtained by solving the eigenvalue problem of $H_\theta$ in Eq.~(\ref{eq:eigen}).
In CSM, we obtain all the energy eigenvalues $E$ of bound and unbound states on a complex energy plane, 
governed by the ABC-theorem\cite{ABC}.
In this theorem, it is proved that the boundary condition of the Gamow resonances 
is transformed to the damping behavior at the asymptotic region.
This condition enables us to use the same theoretical method to obtain the many-body resonances as that for the bound states. 
For a finite value of $\theta$, the Riemann branch cuts are rotated down by $2\theta$,
and continuum states such as of the $^6${He}+$n$ $^5$He+2$n$ and $^4$He+3$n$ channels are obtained on these cuts with the $2\theta$ dependence (See Fig. \ref{fig:ene}). 
On the contrary, bound states and resonances are discrete and obtained independently of $\theta$.
Hence they are located separately from the many-body continuum spectra on the complex energy plane.
We can identify the resonances with complex eigenvalues of $E=E_r-i\Gamma/2$ where $E_r$ and $\Gamma$ are resonance energies measured from the threshold and decay widths, respectively. 
We take the value of $\theta$ as 29$^\circ$ in the present calculation.

\section{Results}

\subsection{Energy spectra of $^6$He and $^7$He}

\begin{figure}[t]
\centering
\includegraphics[width=8.2cm,clip]{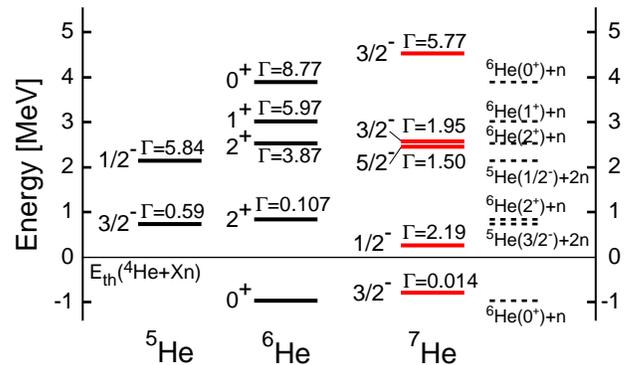}
\caption{(Color online) Energy eigenvalues of the obtained $^{5,6,7}$He resonances measured from the $^4$He+$Xn$ threshold.}
\label{fig:567}
\end{figure}
\begin{figure}[t]
\centering
\includegraphics[width=8.2cm,clip]{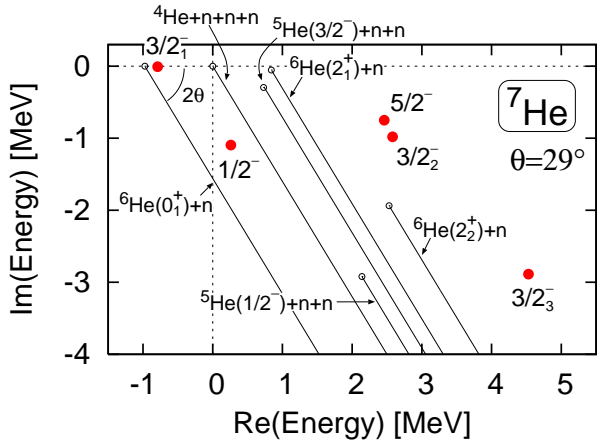}
\caption{(Color online) Energy eigenvalues for the $^7$He resonances (solid circles) in the complex energy plane.
The continuum states rotated down by $2\theta$ are schematically displayed with the cut lines.}
\label{fig:ene}
\end{figure}
\begin{figure}[t]
\centering
\includegraphics[width=8.2cm,clip]{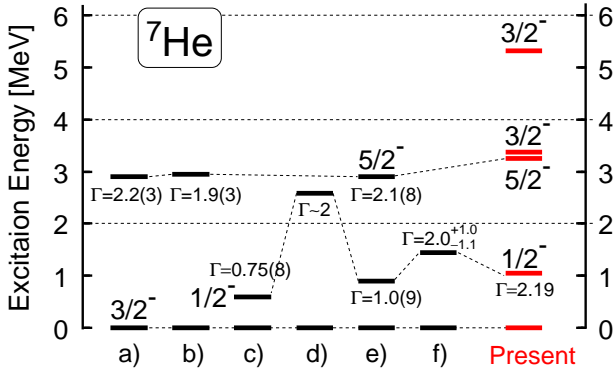}
\caption{(Color online) Excitation spectra of $^7$He in comparison with the experiments
(a)\cite{Ko99}, b)\cite{Bo01}, c)\cite{Me02}, d)\cite{Wu05}, e)\cite{Sk06}, f)\cite{Ry06}).}
\label{fig:exp}
\end{figure}

\begin{table}[t]
\caption{Energy eigenvalues of the $^7$He resonances measured from the $^4$He+3$n$ threshold.
The values with parentheses are the ones fitted to the position of the observed resonance energy of the ground state.}
\label{ene}
\centering
\begin{ruledtabular}
\begin{tabular}{c|cccc}
            & Energy~[MeV]       &  Width~[MeV]  \\ \hline
 $3/2^-_1$  & $-0.790$~~($-$0.54) &  0.014~~(0.14) \\
 $3/2^-_2$  & $2.58$             &  1.95         \\
 $3/2^-_3$  & $4.53$             &  5.77         \\
 $1/2^-  $  & $0.26$             &  2.19         \\
 $5/2^-  $  & $2.46$             &  1.50         \\
\end{tabular}
\end{ruledtabular}
\end{table}

We first discuss the calculational results for the dominant configurations and structures of the $^6$He states, shown in Fig.~\ref{fig:567},
which are useful to understand the $^7$He structures.
For the $^6$He ground state,  the matter radii of 2.36 fm reproduces the experiment (2.33$\pm$0.04 fm)\cite{Ta92}
and the proton and neutron radius are obtained as 1.81 fm and 2.59 fm, respectively.
The dominant configurations are $(p_{3/2})^2$ and $(p_{1/2})^2$  with their squared amplitudes 
of $0.920$ and $0.040$, respectively. 
The contribution of $sd$ shell is $0.039$, which is the same order as the $(p_{1/2})^2$ component.
The dominant configurations of $2^+_1$, $0^+_2$, $2^+_2$ and $1^+$ excited resonant states in $^6$He
are $(p_{3/2})^2_{2^+}$, $(p_{1/2})^2_{0^+}$, $(p_{3/2}p_{1/2})_{2^+}$ and $(p_{3/2}p_{1/2})_{1^+}$ 
with $0.900+i0.010$, $0.967+i0.007$, $0.903+i0.024$ and $0.989-i0.001$, respectively.
Here, it should be noted that an amplitude of a resonance becomes a complex number and its real part has a physical meaning
while an imaginary part has a small value.
These $^6$He states together with a neutron compose the thresholds of $^7$He, and their positions in the complex energy plane
are located at the starting points of the $2\theta$-rotated cuts in the complex scaling method,
as shown in Fig.~\ref{fig:ene}.

\begin{table*}[t]
\centering
\caption{Configurations of valence neutrons with their squared amplitudes $C^2_\beta$ in the $^7$He resonances. $\bar{\ell}_j$ is the orthogonal state of $\ell_j$.}
\label{conf}
\begin{ruledtabular}
\begin{tabular}{ll|ll|lr}
\multicolumn{2}{c|}{$3/2^-_1$}          & \multicolumn{2}{c|}{$3/2^-_2$}                 &\multicolumn{2}{c}{$3/2^-_3$}         \\ \hline 
$(p_{3/2})^3$          &$0.920+i0.0004$ &$(p_{3/2})^2(p_{1/2})$          &$0.883+i0.044$ &$(p_{3/2})(p_{1/2})^2$  & $0.926+i0.161$ \\
$(p_{3/2})(p_{1/2})^2$ &$0.026+i0.004$  &$(p_{3/2})(p_{1/2})^2$          &$0.093-i0.029$ &$(p_{3/2})^2(p_{1/2})$  & $0.117-i0.154$ \\
$(p_{3/2})^2(p_{1/2})$ &$0.016-i0.004$  &$(d_{5/2})(d_{3/2})(p_{3/2})$   &$0.012-i0.013$ &$(d_{3/2})^2(p_{3/2})$  &$-0.031-i0.012$ \\
$(d_{5/2})^2(p_{3/2})$ &$0.015+i0.002$  &$(d_{5/2})^2(p_{1/2})$          &$0.003+i0.001$ &$(p_{3/2})^3$           & $0.007-i0.008$ \\
\hline
sum                    &$0.978+i0.002$  & sum                            &$0.991+i0.002$ &  sum                   & $1.018-i0.013$ \\
\end{tabular}
\vspace*{0.2cm}
\begin{tabular}{lr|lr}
\multicolumn{2}{c|}{$1/2^-$}                    & \multicolumn{2}{c}{$5/2^-$}                       \\ \hline	
$(p_{3/2})^2(p_{1/2})$        & $0.968-i0.097$  & $(p_{3/2})^2(p_{1/2})$         & $0.983-i0.004$   \\	
$(d_{5/2})^2(p_{1/2})$        & $0.022+i0.002$  & $(p_{3/2})^2(\bar{p}_{3/2})$   &$-0.012+i0.004$   \\	
$(p_{1/2})^2(\bar{p}_{1/2})$  & $0.012+i0.021$  & $(1s_{1/2})(d_{5/2})(p_{3/2})$ & $0.008-i0.0004$  \\	
$(1s_{1/2})^2(p_{1/2}) $      &$-0.010+i0.073$  & $(1s_{1/2})(d_{3/2})(p_{3/2})$ & $0.006+i0.003$   \\        
\hline
sum                           & $0.991-i0.002$  & sum                            & $0.984+i0.002$   \\        
\end{tabular}
\end{ruledtabular}
\end{table*}

Next, for the $^7$He resonances, we obtain five states which are all located above the $^6$He(ground state)+$n$ threshold.
We list their energies and decay widths in Table \ref{ene} measured from the $^4$He+3$n$ threshold energy. 
All excited resonant states except for the ground state are obtained above the $^4$He+3$n$ threshold.
In Fig.~\ref{fig:567}, we summarize the energy spectra for $^7$He with those of $^{5,6}$He.
In Fig.~\ref{fig:ene}, we display the energy eigenvalues of the $^7$He resonances together with 
the many-body continuum cuts on the complex energy plane. 
The energy of the ground state is reproduced as $E_r$=0.184 MeV measured from the $^6$He+$n$ threshold.
The result is slightly overbound with respect to the experiments ($E_r=0.44(2)$ MeV\cite{Ko99} and $0.36(5)$\cite{Sk06} MeV).
Due to this overbinding, the decay width is smaller than the experiments of $\Gamma \sim 0.16$ MeV. 
When we fit the above energy of $E_r=0.44$ MeV by reducing the strength of $V^{\alpha n n}$,
the decay width $\Gamma$ becomes 0.14 MeV and nicely agrees with the experiments. The overbinding problem is discussed later.

In Fig.~\ref{fig:exp}, we display the excitation energies in comparison with the various results of the experiments. 
We found the $5/2^-$ state, whose position agrees with the several experiments\cite{Ko99,Bo01,Sk06},
and the obtained decay width of 1.50 MeV is a little smaller than these experiments.
As seen from Fig.~\ref{fig:exp}, the obtained $3/2^-_2$ state is degenerated with the $5/2^-$ state and their decay widths do not differ so much (See Table \ref{ene}).
This result suggests the superposed observation of the two states in this energy region.
We found one broad $1/2^-$ resonance with a low excitation energy of $E_x$=1.05 MeV. 
Three experiments report the $1/2^-$ state with a low excitation energy at around 1 MeV\cite{Me02,Sk06,Ry06},
while the experimental uncertainty is large. Other experiments\cite{Wu05,Bo05} exclude the possibility of the low excitation energy
of this state and instead, suggest the higher excitation energy of $E_x$=2.6 MeV\cite{Wu05}. 
It is desired that further consistent experimental data are coming.

We discuss the structures of each resonance in detail.
In CSM, resonances are precisely described as eigenstates solved using an $L^2$ basis functions, and thus 
have finite amplitudes normalized as unity totally.
We list the main configurations (squared amplitudes $C_\beta^2$ in Eq. (\ref{WF3}) ) for the $^7$He resonances in Table \ref{conf}.  
In general, the square amplitude of each configuration of the resonant states can be a complex number, while the total amplitude of 
the state is normalized to be unity. The physical interpretation of the imaginary parts in physical quantity
of the resonances is still an open problem\cite{Be96,Ho97}. 
However, the amplitudes of the dominant components are almost real values for every resonance, because their imaginary parts are very small. Hence, it is expected that we can discuss the physical meaning 
of the dominant components of the resonances in the same way as the case of bound states.
It is furthermore found that the imaginary parts of the dominant configurations are canceled to each other for every resonance and 
their summations have much smaller imaginary parts.
When we consider all the available configurations, the summations conserve unity due to the normalization of the states.

For the $3/2^-$ ground state, 
our results indicate that the $(p_{3/2})^3$ configuration is dominant with a small mixing of the $p_{1/2}$ component.
For the excited $3/2^-_2$ state, we obtained the interesting result; 
one neutron occupies the $p_{1/2}$ orbit and the residual two neutrons in $p_{3/2}$ forms the spin of $2^+$, 
which corresponds to $^6$He($2^+_1$),
because the first excited $2^+$ state of $^6$He has been studied to have the dominant $(p_{3/2})^2$ configuration\cite{Ao95}.
The importance of the $^6$He($2^+_1$)+$n$ configuration in the $3/2^-_2$ state of $^7$He is also discussed later using $S$-factors.
Two-particle excitation of the $(p_{1/2})^2$ component is mixed by about 9\%.
The other excited $3/2^-_3$ state is dominated by the $(p_{3/2})(p_{1/2})^2$ configuration,
in which the $(p_{1/2})^2$ part is the same configuration of $^6$He($0^+_2$).
From the configurations, the several excited states of $^7$He can be described by the $^6$He+$n$ configuration.
The $^6$He component in $^7$He is shown via $S$-factors in detail later.

The $1/2^-$ state corresponds to the one particle excitation from the ground state.
Its decay width (2.19 MeV) is twice larger than the resonance energy (1.05 MeV).
This property is similar to the $1/2^-$ case of $^5$He in the $^4$He+$n$ system.
In comparison with the $^5$He case, whose resonance energy is 2.13 MeV with the decay width of 5.84 MeV,
the $1/2^-$ state of $^7$He has a smaller excitation energy, and is closer to the threshold of $^6$He+$n$.
The difference comes from the residual two neutrons occupying the $p_{3/2}$ orbit in $^7$He.
The attraction between the $p_{1/2}$ neutron and other two neutrons makes the energy of the $1/2^-$ state lower.

In the $5/2^-$ state, the $2^+$ component of $(p_{3/2})^2$ plus $p_{1/2}$ is a dominant configuration.
This coupling scheme is similar to the $3/2^-_2$ case. 
Furthermore, in every resonance, $1s$ and $0d$ wave configurations are mixed slightly being coupled with the $p$ orbits.

We return to the overbinding problem of the ground state.
Our model reproduces the energies of $^{5,6}$He, and in this sense 
the slight overbinding of $^7$He with respect to the $^4$He+$3n$ threshold suggests the problem of the employed interactions. 
It is interesting to see the contributions of the higher partial waves beyond $\ell=2$ for the valence neutrons 
while tuning the energies of $^{5,6}$He again, although the essential results of the energy spectra and the configuration mixing
would not change. 
On the other hand, the rearrangement of $^4$He inside $^7$He is expected\cite{Cs93,Ar99,My05}, 
which is not included explicitly in the present model.
The tensor correlation produces the strong $2p$-$2h$ excitations in $^4$He, 
which are coupled with the motions of valence neutrons outside $^4$He\cite{My05,My07,My07b,Te60}.
It would be interesting to see these two kinds of effects on the structures not only of the ground state,
but also of the excited states in He isotopes\cite{Ik06}.

\subsection{Spectroscopic factors of $^7$He}

Finally we investigate $S$-factors of the $^6$He-$n$ components for the $^7$He resonances. 
Before proceeding to the results, we would like to mention $S$-factors for Gamow states carefully.
It should be noted that $S$-factors are not necessarily positive definite for Gamow states.
Since Gamow states belong to the eigenstates having complex energies, 
their matrix elements of the physical quantities have complex numbers generally.
$S$-factors for the Gamow states are defined by the squared matrix elements, but not Hermitian products, 
due to the bi-orthogonal properties of the states\cite{Mi07,My01,Be96,Be68} as
\begin{eqnarray}
    S^{J,\nu}_{J',\nu'}
&=& \sum_\alpha S^{J,\nu}_{J',\nu',\alpha} ,
\\
    S^{J,\nu}_{J',\nu',\alpha}
&=& \frac{1}{2J+1} \langle \widetilde{\Psi}^J_\nu(^{7}{\rm He})||a^\dagger_\alpha ||\Psi^{J'}_{\nu'}(^{6}{\rm He}) \rangle^2,
    \label{eq:S}
\end{eqnarray}
where $a^\dagger_\alpha$ is defined in Eq.~(\ref{WF1}).
$J$ and $J'$ are the spins of $^7$He and $^6$He, respectively.
$\nu$ ($\nu'$) is an index to distinguish the obtained eigenstates of $^7$He with $J$ ($^6$He with $J'$)
expressed in Eq.~(\ref{WF3}).
We take a summation over the possible configurations $\alpha$ of a valence neutron.
$\{\widetilde{\Psi}^J_\nu\}$ are bi-orthogonal states of $\{\Psi^J_\nu\}$.
In this expression, $S^{J,\nu}_{J',\nu'}$ are allowed to be complex values and include the physical information 
of the resonant wave functions.
In general, an imaginary part in $S$-factors frequently becomes large relative to the real part for a broad resonance,
which has a large decay width.
When an imaginary part of the matrix element is rather smaller than the real part,
physical interpretation is allowable for the matrix element as a usual $S$-factor, similar to the amplitudes 
of the configurations for the Gamow states as discussed in Table \ref{conf}.
For the obtained resonances, we checked that the real parts of the calculated results are consistent with those obtained in the bound state approximation for resonances.
It is considered that the matrix elements of the Gamow states could be connected to those of the bound states
in the analytical continuation between them by adjusting the strength of the interaction in the Hamiltonian.

The sum rule value for the $S$-factors of Gamow states could be considered, which corresponds to the associated particle number\cite{Ao06,Ho97}.
When we count all the obtained complex $S$-factors for not only Gamow states but also the non-resonant continuum states of the subsystems,
the summed value of the $S$-factors becomes real and satisfies the particular sum rule value derived from the completeness relation of the obtained eigenstates.
In that case, the imaginary part of the summed $S$-factors is automatically canceled out, as similar to the amplitudes of the configurations shown in Table~\ref{conf} and also to the transition strength functions\cite{My03,Ao06}.
In the case of $^7$He with the $^6$He-$n$ decompositions, the summed value of the $S$-factor $S^{J,\nu}_{J',\nu'}$ in Eq.~(\ref{eq:S}) by taking all the $^6$He states is given as
\begin{eqnarray}
    \sum_{J',\nu'}\ S^{J,\nu}_{J',\nu'}
&=& \sum_{\alpha,m}\ 
    \langle \widetilde{\Psi}^{JM}_\nu(^{7}{\rm He})|a^\dagger_{\alpha,m} a_{\alpha,m}| \Psi^{JM}_\nu(^{7}{\rm He}) \rangle
    \nonumber
    \\
&=& 3\ ,
\end{eqnarray}
where we use the completeness relation of $^6$He ($1={\displaystyle\int}\hspace*{-0.4cm}\sum_{J',M',\nu'}| \Psi^{J'M'}_{\nu'}(^{6}{\rm He}) \rangle \langle \widetilde{\Psi}^{J'M'}_{\nu'}(^{6}{\rm He})|)$.
Here $M$ ($M'$) and $m$ are the $z$-components of the wave functions of $^7$He ($^6$He) and of the creation operator
of the valence neurons, respectively.
It is found that the summed value of $S$-factor satisfies the number of valence neutrons of $^7$He for every $^7$He resonance because the state is normalized.

\begin{table}[t]
\caption{Spectroscopic factors of the $^6$He-$n$ components in $^7$He. Details are described in the text.}
\label{sfactor}
\begin{ruledtabular}
\begin{tabular}{c|lll|lll}
            &  \multicolumn{3}{c|}{$^6$He($0^+_1$)-$n$}     &  \multicolumn{3}{c}{$^6$He($2^+_1$)-$n$}      \\
            &  Present      & CK   & VMC   &  Present      & CK   & VMC   \\
\noalign{\hrule height 0.5pt}
$3/2^-_1$   &  $0.75+i0.10$ & 0.59 & 0.53  &  $1.51-i0.40$ & 1.21 & 1.76  \\
$3/2^-_2$   &  $0.03+i0.03$ & 0.06 & 0.06  &  $1.78+i0.06$ & 1.38 & 1.11  \\
$3/2^-_3$   &  $0.01+i0.03$ & ---  & ---   &  $0.02+i0.05$ & ---  & ---   \\ 
$1/2^- $    &  $0.25-i0.47$ & 0.69 & 0.87  &  $0.13-i0.08$ & 0.60 & 0.34  \\
$5/2^- $    &  $0.00+i0.00$ & 0.00 & 0.00  &  $1.37-i0.15$ & 1.36 & 1.20  \\
\end{tabular}

\end{ruledtabular}
\end{table}

In Table \ref{sfactor}, we list the results of $S$-factors for the $^7$He resonances, 
which are calculated using the complex-scaled wave functions and independent of the scaling angle $\theta$.
In our calculation, we also describe $^6$He($2^+_1$) as a Gamow state.
For reference, the results of the conventional Cohen-Kurath shell model (CK) and of the variational Monte Calro (VMC) calculations\cite{Wu05,Pi04} are also shown
with real values due to the bound state approximation for the description of resonances.
The trend seen in our results is roughly similar to the CK and VMC results.
For the $3/2^-_1$ state, the mixing of $^6$He($2^+_1$) component is almost twice of that of the $^6$He($0^+_1$) case.
For the $3/2^-_2$ state, $^6$He($2^+_1$) is strongly mixed from the dominant amplitude of $(p_{3/2})^2_{2^+}\otimes(p_{1/2})$.
For the $3/2^-_3$ state, the $0^+_1$ and $2^+_1$ states of $^6$He are hardly included because of the $(p_{3/2})\otimes(p_{1/2})^2$ configuration.
Instead of the above two $^6$He states, the $0^+_2$ ($(p_{1/2})^2$) and $2^+_2$ ($(p_{3/2})(p_{1/2})$) states 
of $^6$He may give large contributions for this state\cite{My01}.
For the $1/2^-$ state, even if this state is dominated by a $(p_{3/2})^2\otimes(p_{1/2})$ component, 
the $S$-factor for $^6$He($0^+_1$) is not large.
This indicates that the spatial property of the $(p_{3/2})^2$ component is changed in the $1/2^-$ state of $^7$He
from the halo structure of the neutrons in $^6$He($0^+_1$).
This is because that the $1/2^-$ state is located above the $^4$He+3$n$ threshold and can decay to four particles.
In fact, when we locate this state just below 0.5 MeV from the $^4$He+3$n$ threshold energy by adjusting interaction, 
the $S$-factor becomes $0.79-i0.35$ and its real part gets close to unity.
The $^6$He($2^+_1$) component is small in this state. 
The $1/2^-$ state also shows the large imaginary part of the $S$-factor, which comes
from the large decay width of this state.
The present $S$-factors correspond to the components of $^6$He in the $^7$He resonances, similar to 
the results shown in Table \ref{conf}.
However, it is still difficult to derive the definite conclusion of the interpretation of this imaginary part at this stage.
The further theoretical development and analysis would be desired to solve this problem.
For the $5/2^-$ state, the $^6$He($2^+_1$) component is included well.
For the summary of the results of the $S$-factors, the obtained $^7$He states are not considered 
to be purely single particle states coupled with the $^6$He ground state. 
The excitation of $^6$He into $2^+_1$ is important in several states.
\section{Summary}
We have investigated the resonance structures of $^7$He with the cluster orbital shell model.
The boundary condition for many-body resonances is accurately treated in the complex scaling method. 
The decay thresholds concerned with subsystems are described consistently.
As a result, we found five resonances, which are dominantly described by the $p$-shell configurations 
and the small contributions come from the $sd$-shell.
The $1/2^-$ state is predicted in a low excitation energy region with a large decay width.
We further investigate the spectroscopic factor of the $^6$He-$n$ component.
It is found that the $^6$He(2$^+_1$) state contributes largely in the ground and the several excited states of $^7$He.


\begin{acknowledgments}
The authors would like to thank Prof. H. Masui for valuable discussions. 
This work was performed as a part of the ``Research Project for Study of
Unstable Nuclei from Nuclear Cluster Aspects (SUNNCA)'' at RIKEN and
supported by a Grant-in-Aid from the Japan Society for the Promotion of Science (JSPS, No.18-8665).
Numerical calculations were performed on the computer system at RCNP.
\end{acknowledgments}

\end{document}